\documentclass{article}[10pt]
\usepackage[dvips]{graphicx}
\usepackage{amsmath}
\usepackage{amssymb}
\usepackage{epsfig}
\begin{document}


%
%

\title{CURVATURE AND TOPOLOGICAL EFFECTS ON DYNAMICAL SYMMETRY 
BREAKING IN A FOUR- AND EIGHT-FERMION INTERACTION MODEL}

\author{Masako Hayashi\\
Department of Physics, Hiroshima University,\\
Higashi-Hiroshima 739-8526, Japan\\[2mm]
Tomohiro Inagaki\\
Information Media Center, Hiroshima University,\\
Higashi-Hiroshima 739-8521, Japan
}




\maketitle


\begin{abstract}
A dynamical mechanism for symmetry breaking is investigated
under the circumstances with the finite curvature, finite size and 
non-trivial topology. A four- and eight-fermion interaction 
model is considered as a prototype model which induces symmetry 
breaking at GUT era. Evaluating the effective potential in 
the leading order of the $1/N$-expansion by using the 
dimensional regularization, we explicitly calculate the phase
boundary which divides the symmetric and the broken phase
in a weakly curved space-time and a flat space-time with 
non-trivial topology, $R^{D-1} \otimes S^1$.
\end{abstract}


\section{Introduction}
It is believed that a fundamental theory with a higher symmetry
is realized at high energy scale. In a grand unified theory, the
higher gauge symmetry is broken down to the standard model gauge 
symmetry, $SU_c(3)\otimes SU(2)_L\otimes U(1)_Y$.
The evolution of the universe has played an important role for 
the symmetry breaking. On the other hand the symmetry breaking 
affects the space-time structure of the universe. Therefore it 
is expected that some evidences of symmetry breaking at GUT era 
will be observed in the structure of the current universe.

One of the well-known mechanism of the symmetry breaking is found
for the chiral symmetry breaking in QCD. Below the QCD scale a 
composite operator which is constructed by quark and anti-quark 
fields develops a non-vanishing expectation value and the quark 
fields acquire mass term. Then the chiral symmetry is spontaneously 
broken. It is caused by non-perturbative effect of QCD dynamics 
and called as dynamical symmetry breaking. This mechanism can 
be applied to strong coupling gauge theories at GUT scale.

Thus we have launched our plan for a systematic study on the 
dynamical symmetry breaking in extreme conditions at the early
universe. A simple four-fermion interaction model are often used 
for the analysis of the dynamical symmetry breaking. The curvature 
effects has been studied in two\cite{Itoyama,BK}, three\cite{EOS}, 
four\cite{IMO} and arbitrary dimensions.\cite{Inagaki}. If the
background space-time has a positive curvature, the symmetry breaking 
is suppressed. The broken symmetry is restored for a large positive
curvature, while only the broken phase can be realized in a 
space-time with a negative curvature.\cite{Gorbar:1999wa}
The finite size effect has been investigated in the cylindrical
universe,\cite{Vshivtsev:1995fh} and the torus 
universe.\cite{Kim:1987db,Song:1990dm,Kim:1994es,IIYF,Abreu:2006pt}
The finite size effect restore the broken symmetry if we adopt the 
anti-periodic boundary condition to the fermion fields. The fermion 
fields with the periodic boundary condition breaks the chiral 
symmetry. A combined effect of the space-time curvature and the 
topology has been also discussed in a weakly curved 
space-time,\cite{ELO2} the maximally symmetric 
space-time\cite{IMM,ELO} and Einstein space\cite{IIM,Ebert:2008tp}. 
For a review, see for example Ref.~\cite{IMO2}.

Strong gauge interactions at high energy scale can be represented 
by various operators in low energy effective models. It is not 
always valid to neglect higher dimensional operators.
't Hooft introduced a determinantal interactions 
in a low energy effective model of QCD to deal with the explicit 
breaking of the $U_A(1)$ symmetry.\cite{tHooft}
R. Alkofer and I. Zahed considered an eight-fermion interaction 
to explain the pseudoscalar nonet mass spectrum.\cite{Alkofer:1990uh}
An influence of higher derivatives is also considered in Ref.
\cite{ELO3}.

In the present paper we consider four- and eight-fermion 
interactions and investigate the phase structure of the model
in a weakly curved space-time and a space-time with non-trivial 
topology. In Sec.~2 we introduce a four- and eight-fermion interaction
model and calculate the effective potential in the leading order of
the $1/N$ expansion. In Sec.~3 we discuss the curvature effect on
the dynamical symmetry breaking. It is assumed that the space-time 
curves slowly. We keep only terms independent of the space-time 
curvature $R$ and linear to $R$ and study the curvature induced 
phase transition. We consider a cylindrical universe, 
$R^{D-1} \otimes S^1$ in Sec.~4.
It is a flat space-time with non-trivial topology. We adopt a 
periodic and an anti-periodic boundary conditions for the fermion 
fields. The phase structure of the model is studied as the size 
for $S^1$ direction varies for each boundary conditions.
In Sec.~5 we give some concluding remarks.

\section{Four- and Eight-Fermion Interaction Model}
The space-time structure modifies the phase structure of the 
four-fermion interaction model.\cite{IMO2} 
The simplest model is composed of $N$-flavor fermions with a scalar
type four-fermion interaction.
In an extreme conditions some interactions described by higher 
dimensional operators may have nonnegligible contribution to the 
phase structure. 
Thus we extend the model with higher dimensional operators to find a 
general and an essential feature for the dynamical symmetry breaking.
We employ scalar type four- and eight-fermion interactions in a 
curved space-time and start from the action,
\begin{eqnarray}
S&=&\int d^D x \sqrt{-g}\left[\sum_{i=1}^{N}
 \bar{\psi}_i(x) i\gamma^\mu(x)\nabla_\mu\psi_i(x)
 + \frac{G_1}{N} (\sum_{i=1}^{N} \bar{\psi}_i(x)\psi_i(x))^2 \right.
\nonumber\\
&&\left. + \frac{G_2}{N^3} 
(\sum_{i=1}^{N} \bar{\psi}_i(x)\psi_i(x))^4 \right],
\label{action:1}
\end{eqnarray}
where the index $i$ represents the flavors of the fermion field 
$\psi$, $N$ is the number of flavors, $g$ the determinant of the 
metric tensor, $\gamma^{\mu}(x)$ the Dirac matrix in curved 
space-time and $\nabla_\mu\psi$ the covariant derivative for the 
fermion field $\psi$. The coupling constants for the four- and the 
eight-fermion interactions, $G_1$ and $G_2$, should be fixed 
phenomenologically. Since the model is unrenormalizable, we need
one more parameter to regularize radiative corrections.
In this paper we adopt the dimensional regularization for fermion
loop corrections. We leave the space-time dimension, $D$, as an 
arbitrary parameter to be fixed 
phenomenologically.\cite{Naka,He,IKM,JR,IKK}

The action (\ref{action:1}) is a simple extension of the Gross-Neveu
model\cite{GN} with the same symmetry. It is invariant under the 
discrete chiral transformation,
\begin{equation}
\bar{\psi}_i\psi_i \rightarrow -\bar{\psi}_i\psi_i,
\label{chiral}
\end{equation}
and the global flavor transformation,
\begin{equation}
\psi_i \rightarrow 
\left(\exp (i\sum_a \theta_a T_a)\right)_{ij} \psi_j,
\label{flavor}
\end{equation}
where $T_a$ are generators of the flavor $SU(N)$ symmetry. The flavor
symmetry allows us to work in a scheme of the $1/N$ expansion.
Below we neglect the flavor index.

The generating functional of the model is given by
\begin{equation}
Z=\int {\cal D}\psi{\cal D}\bar{\psi}e^{iS}.
\label{gen1}
\end{equation}
It should be noted that we set the path integral measure to keep the 
general covariance. For practical calculations it is more convenient 
to introduce the auxiliary fields.
We consider a constant integral,
\begin{equation}
C=\int D\sigma \delta\left(\sigma+\frac{2G_1}{N}\bar{\psi}\psi\right) .
\label{Def:aux}
\end{equation}
The delta function in Eq.(\ref{Def:aux}) is described by the integral 
form,
\begin{equation}
C=C'\int D\sigma Ds e^{i S_a} ,
\end{equation}
where $S_a$ is given by
\begin{equation}
S_a = i\int \sqrt{-g} d^D x
\left[-\frac{N}{2G_1}s\left(\sigma+\frac{2G_1}{N}\bar{\psi}\psi\right)\right].
\end{equation}
Since $C$ is a constant, we are free to insert it on the right-hand side
in Eq.(\ref{gen1}). The fermion bilinear, $\bar{\psi}\psi$, is 
replaced by the auxiliary field $\sigma$. Thus the generating 
functional, $Z$, is rewritten as
\begin{equation}
Z=\int D\psi D\bar{\psi} D\sigma Ds e^{iS_y},
\end{equation}
where $S_y$ is given by
\begin{eqnarray}
S_y&=&S+S_a
\nonumber \\
&=&\int d^Dx \sqrt{-g}\left[\bar{\psi}
(i\gamma^\mu(x)\nabla_\mu -s) \psi 
+ \frac{N }{4G_1}\sigma^{2}
- \frac{N}{2G_1}s\sigma 
+ \frac{NG_2 }{16G_1^{4}}\sigma^{4}
\right].
\label{Sy1}
\end{eqnarray}
The eight-fermion interactions in the original action are replaced by 
the auxiliary fields, $s$ and $\sigma$.

From the action (\ref{Sy1}) we obtain the equations of motion 
for the auxiliary fields,
\begin{equation}
\sigma=-\frac{2G_1}{N}(\bar{\psi}\psi),
\label{EOM1}
\end{equation}
and
\begin{equation}
s=\sigma+\frac{G_2}{2G_1^3}\sigma^3. 
\label{EOM2}
\end{equation}
If a non-vanishing expectation value is assigned for the 
auxiliary field, $\sigma$, the composite operator, $\bar{\psi}\psi$, 
also has a non-vanishing expectation value and the chiral symmetry
is eventually broken. 
A non-vanishing expectation value for the auxiliary field, $s$, 
generates the fermion mass term.
Substituting the equations (\ref{EOM1}) and (\ref{EOM2}) 
into the action (\ref{Sy1}), we can reproduce the original 
one (\ref{action:1}).

For later convenience we shift the field $\sigma$ to 
$\sigma' \equiv \sigma-s$ and diagonalize the mass term 
for the auxiliary fields. Thus the action (\ref{Sy1}) reads
\begin{equation}
S_y
=\int d^Dx \sqrt{-g}\left[\bar{\psi}
(i\gamma^\mu(x)\nabla_\mu -s) \psi 
+\frac{N}{4G_1}\sigma'^2-\frac{N}{4G_1}s^2
+\frac{G_2N}{16G_1^4}(\sigma'+s)^4
\right] .
\label{Sy2}
\end{equation}
To include quantum corrections we calculate the effective 
action. Performing the path-integral over the fermion fields, we 
calculate the effective action $\Gamma[s, \sigma']$ in the leading 
order of the $1/N$ expansion,
\begin{eqnarray}
\Gamma[s(x), \sigma'(x)]&=&\int d^Dx \sqrt{-g}\left[
\frac{N}{4G_1}\sigma'^2-\frac{N}{4G_1}s^2
+\frac{NG_2}{16G_1^4}(\sigma'+s)^4\right]
\nonumber \\
&&+i\mbox{Tr}\int_0^sdmS(x,x;m),
\label{effectiveA}
\end{eqnarray}
where Tr represents the trace over the flavor, spinor and space-time 
coordinates. We want to find a ground state of the system described 
by the action (\ref{Sy2}). We assume that the ground state is static 
and homogeneous and put $s(x)=s$ and $\sigma(x)=\sigma$ 
constants independent of the space-time coordinate $x$.
Then we can obtain the effective potential,
\begin{equation}
V(s, \sigma')=-\frac{1}{4G_1}\sigma'^2+\frac{1}{4G_1}s^2
-\frac{G_2}{16G_1^4}(\sigma'+s)^4-i\mbox{tr}\int_0^s dm S(x,x;m),
\label{pot}
\end{equation}
where tr shows the trace over only the spinor and we omit the over 
all factor $N$. $S(x,y;m)$ is the spinor two-point function which 
satisfies the Dirac equation in curved space-time,
\begin{equation}
(i\gamma^\mu\nabla_\mu-m)S(x,y;m)=\frac{1}{\sqrt{-g}}\delta^D(x-y).
\label{eq:2p}
\end{equation}

The effective potential (\ref{pot}) gives the energy 
density induced by the fermion fields. The ground state have
to minimize it. Thus we can find the state by the stationary 
condition of the effective potential,
\begin{equation}
\left. \frac{\partial V(s,\sigma')}{\partial s}\right|_{\sigma'}=
\frac{s}{2G_1}-\frac{G_2}{4G_1^4}(\sigma'+s)^3-i\mbox{tr}S(x,y;s)=0
\label{gap:1}
\end{equation}
and
\begin{equation}
\left. \frac{\partial V(s,\sigma')}{\partial \sigma'}\right|_{s}=
-\frac{\sigma'}{2G_1}-\frac{G_2}{4G_1^4}(\sigma'+s)^3=0. 
\label{gap:2}
\end{equation}
Eqs. (\ref{gap:1}) and (\ref{gap:2}) give necessary
conditions for the minimum. From Eq.(\ref{gap:2}) we get
\begin{equation}
s=\left(-\frac{2G_1^3}{G_2}\sigma'\right)^{1/3}-\sigma'.
\label{gapsol:1p}
\end{equation}
In terms of the auxiliary field, $\sigma = s + \sigma'$, 
the equation reads
\begin{equation}
s=\sigma + \frac{G_2}{2G_1^3}\sigma^3.
\label{gapsol:1}
\end{equation}
This equations give the relationship between both 
the auxiliary fields at the stationary point.
Thus the dynamical fermion mass, $m_{d} \equiv \langle s \rangle$
is given by a function of 
$\langle \sigma \rangle \propto \langle \bar{\psi}\psi \rangle$.
Substituting Eq.~(\ref{gapsol:1}) into Eq.~(\ref{gap:1}), we 
obtain
\begin{equation}
s=i2G_1\mbox{tr}S(x,x;s)-4iG_2\left[\mbox{tr}S(x,x;s)\right]^3.
\label{gapsol:2}
\end{equation}
The dynamically generated fermion mass, $m_{d}$, should 
satisfy this gap equation. 
In the case of the second order phase transition
the dynamical fermion mass smoothly disappears at the 
critical point. We can calculate it to take the limit 
$s\rightarrow 0$ for the non-trivial solution of 
Eq.~(\ref{gapsol:2}). Because of the chiral symmetry 
tr$S(x,x;s)$ is proportional to $s$. Thus the second term
in the right-hand side in Eq.~(\ref{gapsol:2}) drops
much faster than the other terms. It shows that the
eight-fermion interaction $G_2$ has nothing to do with
the critical point for the second order phase transition.

We can obtain the self-consistent equation (\ref{gapsol:2}) 
for the dynamically generated fermion mass in other
procedures. An alternative way is possible to cancel
out multi-fermion interactions by introducing the auxiliary 
fields. In Ref.~\cite{HIT} the action (\ref{action:1}) is 
rewritten by the auxiliary fields $\sigma_1$ and $\sigma_2$,
\begin{equation}
S_y'
=\int d^Dx \sqrt{-g}\left[
\bar{\psi}(i\gamma^\mu(x)\nabla_\mu-\sigma_f)\psi
-\frac{N\sigma_1^2}{4G_1}-\frac{N\sigma_2^2}{4G_2}\right], 
\end{equation}
where $\sigma_f$ is given by 
\begin{equation}
\sigma_f \equiv \sigma_1\sqrt{1-\frac{\sigma_2}{G_1}}. 
\end{equation}
In terms of the auxiliary fields $\sigma_1$ and $\sigma_2$
the effective potential is found to be
\begin{equation}
V(\sigma_1, \sigma_2)=\frac{1}{4G_1}\sigma_1^2
+\frac{1}{4G_2}\sigma_2^2
-i\mbox{tr}\int_0^{\sigma_f} dmS(x,x;m).
\label{pot:pre}
\end{equation}
From the stationary condition of the effective 
potential (\ref{pot:pre}) we obtain the self-consistent
equation for the dynamically generated fermion mass,
\begin{equation}
\sigma_f=i2G_1\mbox{tr}S(x,x;\sigma_f)
-4iG_2\left[\mbox{tr}S(x,x;\sigma_f)\right]^3.
\label{gapsol:3}
\end{equation}
It is equivalent to the gap equation (\ref{gapsol:2}).

We can also derive the fermion mass from the original action
(\ref{action:1}). 
In the leading order of the $1/N$ expansion it is given by the
following diagrams,
\begin{eqnarray}
  m_d&=&
  \mbox{
    \begin{minipage}{20mm}
      \hspace*{1mm}
      \includegraphics[height=8mm]{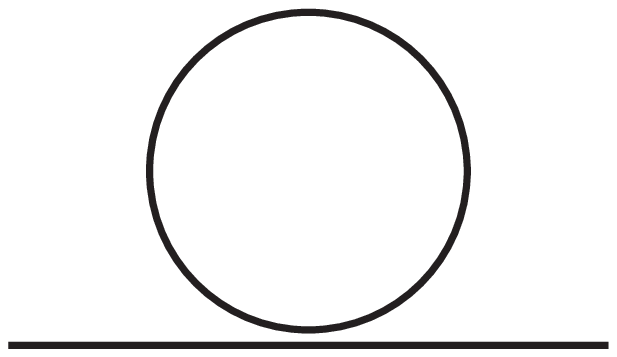}
    \end{minipage}
  }
  +
  \mbox{
    \begin{minipage}{20mm}
      \hspace*{1mm}
      \includegraphics[height=13mm]{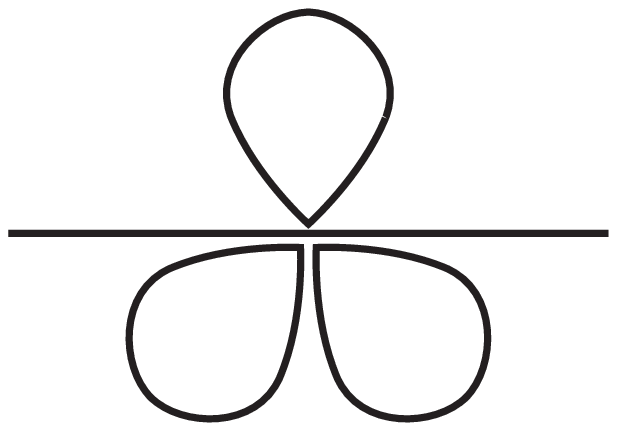}
    \end{minipage}
  }
  \nonumber \\
  &=&2iG_1\mbox{tr}S(x,x;m_d)
  -4iG_2\left(\mbox{tr}S(x,x;m_d)\right)^3.
\end{eqnarray}
Therefore the gap equation (\ref{gapsol:2}) is reproduced again.

Before we discuss the curvature and the topological effect, it
will be instructive to give the type behavior of the effective 
potential Eq. (\ref{pot}) in Minkowski space-time, $R^D$.
The two-point function (\ref{eq:2p}) in $R^D$ is given by
\begin{equation}
S(x,y;s)=\int \frac{d^D k}{(2\pi)^D}e^{-ik(x-y)}
\frac{1}{\gamma^\mu k_\mu -s} .
\label{eq:2p2}
\end{equation}
Inserting Eq.(\ref{eq:2p2}) into Eq.(\ref{pot}), we obtain the
explicit expression for the effective potential,
\begin{eqnarray}
V_0(s, \sigma')&=&-\frac{1}{4G_1}\sigma'^2+\frac{1}{4G_1}s^2
-\frac{G_2}{16G_1^4}(\sigma'+s)^4
\nonumber\\
&&-\frac{\mbox{tr1}}{(4\pi)^{D/2}D}
\Gamma\left(1-\frac{D}{2}\right)|s|^{D}.
\label{v0}
\end{eqnarray}
At the stationary point, the effective potential is 
rewritten as
\begin{eqnarray}
V_0(\sigma)&=&
\frac{1}{4G_1}\left(\sigma^2+
\frac{3G_2}{4G_1^3}\sigma^4\right)
\nonumber\\
&&-\frac{\mbox{tr1}}{(4\pi)^{D/2}D}
\Gamma\left(1-\frac{D}{2}\right)
\left|\sigma+
\frac{G_2}{2G_1^3}\sigma^3\right|^D.  
\label{pot:M}
\end{eqnarray}

In $R^D$ the gap equation (\ref{gapsol:2}) is given by
\begin{equation}
s=2G_1\frac{\mbox{tr}1}{(4\pi)^{D/2}}
\Gamma\left(1-\frac{D}{2}\right)s|s|^{D-2}
+4G_2\left[\frac{\mbox{tr}1}{(4\pi)^{D/2}}
\Gamma\left(1-\frac{D}{2}\right)s|s|^{D-2}\right]^3 .
\end{equation}
For $G_2=0$ we can analytically solve this gap equation
and find a non-trivial solution for a negative $G_1$ at
\begin{equation}
s=\sigma=m_0\equiv \left[-
\frac{(4\pi)^{D/2}}{2|G_1|\mbox{tr}1\Gamma\left(1-D/2\right)}
\right]^{1/(D-2)} .
\label{m0}
\end{equation}
This solution gives the dynamically generated fermion mass for 
$G_2=0$ in $R^D$. Below we normalize all the mass scales by $m_0$ 
for $G_1<0$. In the case of a positive $G_1$ we also use $m_0$ 
to normalize parameters with mass dimension.
Thus the effective potential (\ref{pot:M}) depends on 
$\sigma$, $D$, a sign of $G_1$ and 
$G_r\equiv G_2m_0^2/G_1^3$, 
\begin{eqnarray}
 \frac{V_0(\sigma)}{m_0^D}&=&-sgn(G_1)\frac{\mbox{tr}1}{(4\pi)^{D/2}}
 \Gamma\left(1-\frac{D}{2}\right)
 \left[\left(\frac{1}{2}\frac{\sigma^2}{m_0^2}
   +\frac{3G_r}{8}\frac{\sigma^4}{m_0^4}\right) \right.\nonumber\\
   &&\left.-\frac{1}{D}\left|\frac{\sigma}{m_0}
   +\frac{G_r}{2}\frac{\sigma^3}{m_0^3}\right|^D\right] . 
\end{eqnarray}
In this paper we consider only the four- and the eight-fermion 
interaction. It is expected that contributions from higher 
dimensional operators with mass dimension, $d$, are almost
proportional to $\sigma^{d/3}$. It is not valid to
neglect the contributions for a large $\sigma$.
Thus we restrict our analysis in 
$\langle \sigma \rangle /m_0 \lesssim 1$.

\begin{figure}[tp]
 \begin{minipage}{0.48\hsize}
  \begin{center}
   \includegraphics[width=60mm]{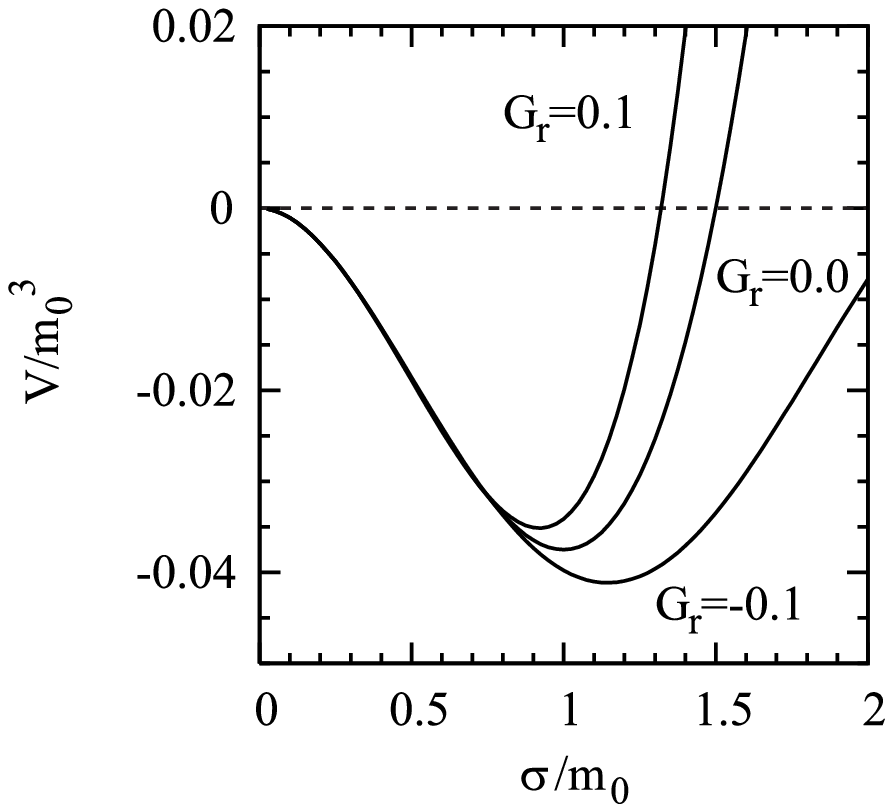}
  \end{center}
  \caption{Behavior of the effective potential in Minkowski 
    space-time for $G_1<0$, $G_r=-0.1, 0, 0.1$ and $D=3$.}
  \label{fig:1}
\end{minipage}\hspace{0.03\hsize}\begin{minipage}{0.46\hsize}
    \begin{center}
      \includegraphics[width=60mm]{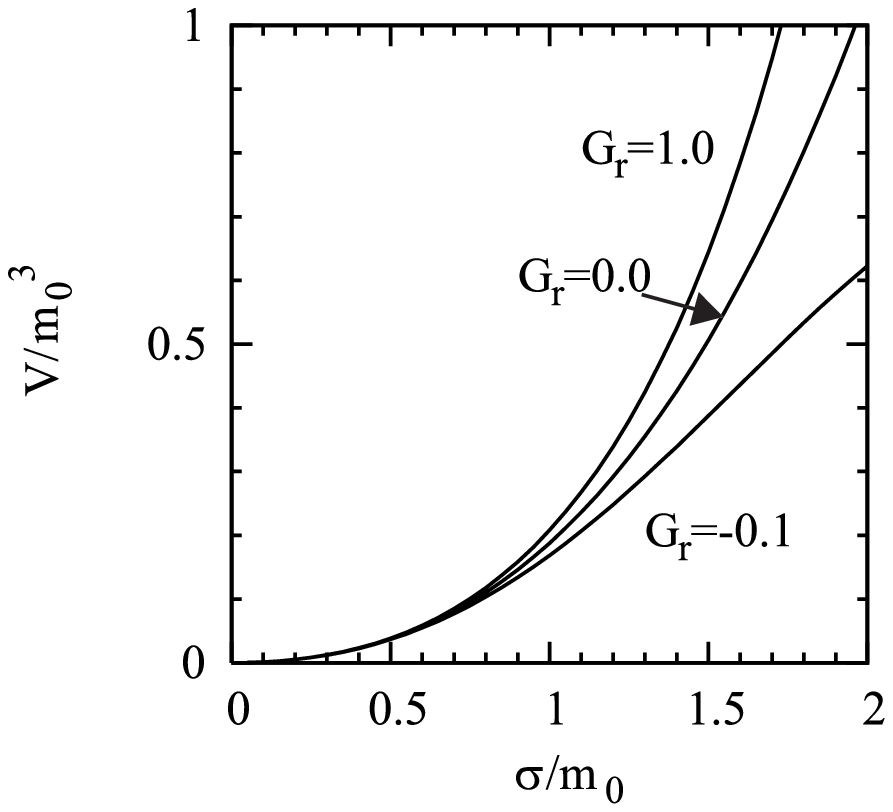}
    \end{center}
    \caption{Behavior of the effective potential in Minkowski 
    space-time for $G_1>0$, $G_r=-0.1, 0, 0.1$ and 
    $D=3$.}
    \label{fig:2}
  \end{minipage}
\end{figure}
Numerically evaluating the effective potential (\ref{pot:M})
as $G_r (\equiv G_2m_0^2/G_1^3)$ varies, we find the phase structure.
We set $D=3$ and show the typical behavior of the effective 
potential in Figs.~\ref{fig:1} and \ref{fig:2} for a negative 
$G_1$ and a positive $G_1$, respectively.
As is shown in Fig. \ref{fig:1}, a positive eight-fermion
coupling, $G_2$, enhances the chiral symmetry breaking, while
a negative one suppresses it.
For a positive $G_1$ we observe only the symmetric phase in 
Fig.~\ref{fig:2}.
It should be noted that we can find a minimum of the 
effective potential, if we evaluate it for a larger 
$\sigma$. 
In the case of $G_r=-0.1$ with a positive 
$G_1$, we find a minimum at $\sigma/m_0 \sim 3.8$.
However, the expectation value of $\sigma$
is too large to keep validity of the model.
It is outside the scope of our interest.

\section{Curvature Induced Phase Transition}
The space-time curvature may play an important role for a phase
transition at the early universe. Here we evaluate the four- and 
eight-fermion interaction model (\ref{action:1}) in a curved 
space-time. Some assumption for the back ground metric is 
necessary to calculate the ground state of the model. Here we
assume that the space-time curves slowly and keep only terms 
independent of the curvature $R$ and terms linear in $R$. 

For this purpose we introduce the Riemann normal coordinate 
system in which the affine connection vanishes at least 
locally\cite{Petrov}. Near the origin $x_0$ the back-ground 
metric is expanded to be
\begin{equation}
g_{\mu\nu}(x)=\eta_{\mu\nu}+\frac{1}{3}R_{\mu\alpha\nu\beta}
(x-x_0)^\alpha(x-x_0)^\beta+O(R_{;\mu},R^2),
\end{equation}
where $\eta_{\mu\nu}$ is the metric tensor on $R^D$, 
diag$(1, -1, \cdots, -1)$ and $R_{\mu\alpha\nu\beta}$ the curvature
tensor at the origin. In the Riemann normal coordinate system
the spinor two-point function (\ref{eq:2p}) is found to 
be\cite{IMO,IMO2}
\begin{eqnarray}
S(x_0,y;s) &=& \int\frac{d^D k}{(2\pi)^D}e^{-ik(x_0-y)}\left[
\frac{\gamma^a k_a +s}{k^2-s^2} - \frac{1}{12}R
\frac{\gamma^a k_a +s}{(k^2-s^2)^2}
\right.
\nonumber \\
&&\left.
+\frac{2}{3}R^{\mu\nu}k_{\mu}k_{\nu}
\frac{\gamma^a k_a +s}{(k^2-s^2)^3}
+\frac{1}{4}\gamma^a\sigma^{cd}R_{cda\mu}k^{\mu}
\frac{1}{(k^2-s^2)^2}
\right]
\nonumber \\
&&+O(R_{;\mu},R^2).
\label{eq:2pR}
\end{eqnarray}

Substituting Eq.(\ref{eq:2pR}) into Eq.(\ref{pot}), we expand
the effective potential in terms of the space-time curvature $R$,
\begin{equation}
V(s,\sigma')=
V_0(s,\sigma')-\frac{\mbox{tr1}}{(4\pi)^{D/2}}\frac{R}{24} 
\Gamma\left(1-\frac{D}{2}\right)|s|^{D-2} +O(R_{;\mu},R^2),
\label{pot:R0}
\end{equation}
where $V_0(s,\sigma')$ is the effective potential (\ref{v0}) for $R=0$.
Since the term linear in $R$ is independent on $\sigma'$, the 
stationary condition (\ref{gap:2}) is not modified. The auxiliary
fields $s$ and $\sigma'$ is rewritten by $\sigma$ by
Eq.(\ref{gapsol:1}) with Eq.{def:tsig} at the stationary point.
Thus the effective potential (\ref{pot:R0}) reads
\begin{equation}
V(\sigma)=V_0(\sigma)
-\frac{\mbox{tr1}}{(4\pi)^{D/2}}\frac{R}{24}\Gamma\left(1-\frac{D}{2}\right)
\left|\sigma+
\frac{G_2}{2G_1^3}\sigma^3\right|^{D-2} +O(R_{;\mu},R^2).
\label{pot:R} 
\end{equation}
The gap equation defined by (\ref{gapsol:2}) reads
\begin{eqnarray}
s&=&2G_1\frac{\mbox{tr}1}{(4\pi)^{D/2}}
\Gamma\left(1-\frac{D}{2}\right)s
\left(|s|^{D-2}-\frac{D-2}{24}R|s|^{D-4}\right)
\nonumber \\
&&+4G_2\left[\frac{\mbox{tr}1}{(4\pi)^{D/2}}
\Gamma\left(1-\frac{D}{2}\right)s\left(|s|^{D-2}
-\frac{D-2}{24}R|s|^{D-4}\right)\right]^3 .
\label{order:R}
\end{eqnarray}
It gives the necessary condition for the dynamically 
generated fermion mass in a weakly curved space-time.
Evaluating the effective potential (\ref{pot:R}) and the
gap equation (\ref{order:R}) in weakly curved space-time,
we study the curvature induce phase transition below.

\subsection{Phase structure for a negative $G_1$}
\begin{figure}[bp]
\begin{minipage}{0.48\hsize}
  \begin{center}
   \includegraphics[width=60mm]{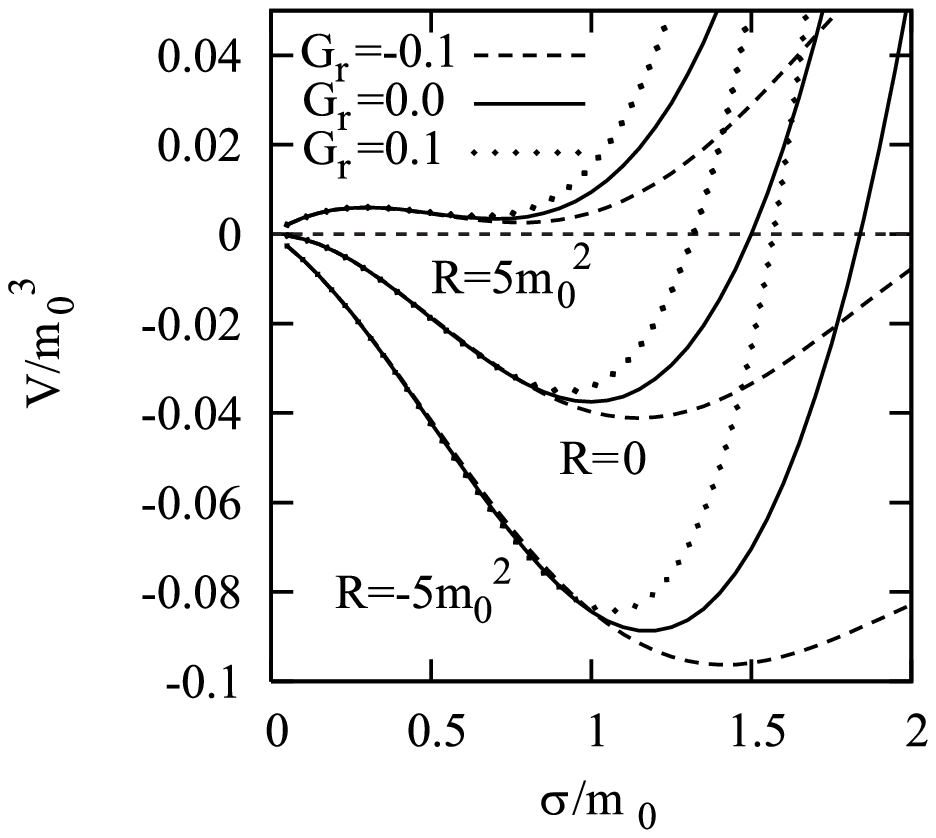}
  \end{center}
  \caption{Behavior of the effective potential for $D=3$,
     $G_1<0$ and $G_r=-0.1, 0, 0.1$ as the curvature, $R$, varies. }
  \label{fig:3}
\end{minipage}\hspace{0.03\hsize}
\begin{minipage}{0.48\hsize}
    \begin{center}
      \includegraphics[width=56mm]{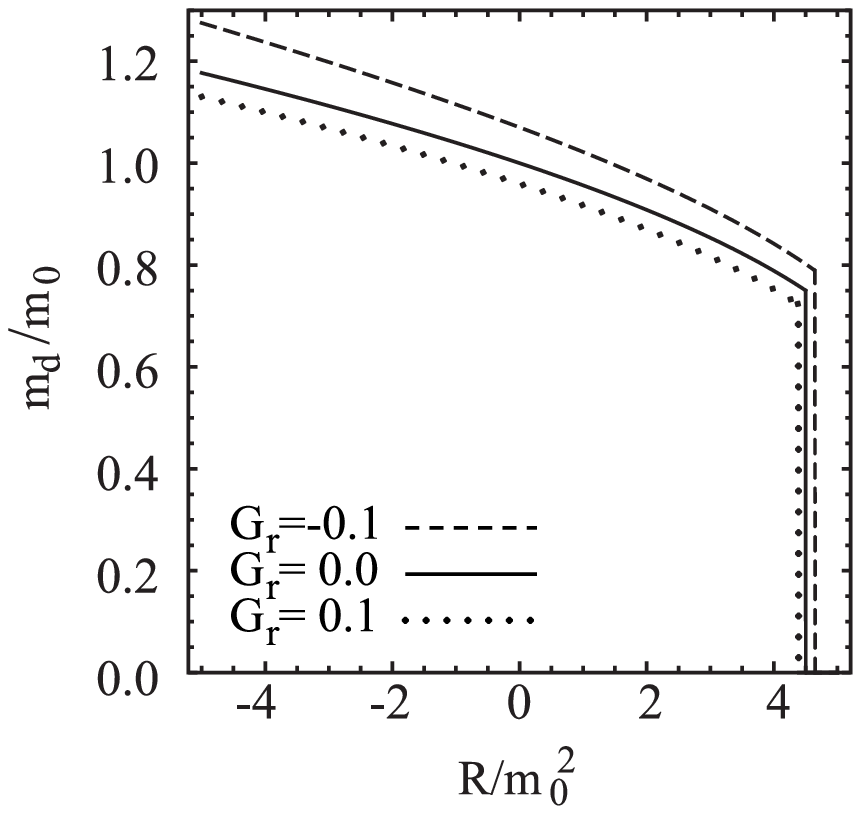}
    \end{center}
    \caption{Dynamically generated fermion mass for $D=3$,
     $G_1<0$ and $G_r=-0.1, 0, 0.1$ as a function of $R$. }
    \label{fig:5}
  \end{minipage}
\end{figure}

For a negative $G_1$ the chiral symmetry is broken in Mikowski
space-time.
We set $D=3$ and numerically draw the typical behavior of the 
effective potential for $G_1<0$ in Fig.~\ref{fig:3}.
A positive $G_r$ suppresses the chiral symmetry breaking, while 
a negative one enhances it.
The chiral symmetry is broken in Minkowski space-time for a negative 
$G_1$. The first order phase transition takes place, as the 
curvature increases. Thus the broken chiral symmetry is restored for 
a large positive $R$. 

\begin{figure}[tbp]
  \begin{minipage}{0.48\hsize}  
  \begin{center}
    \includegraphics[width=58mm]{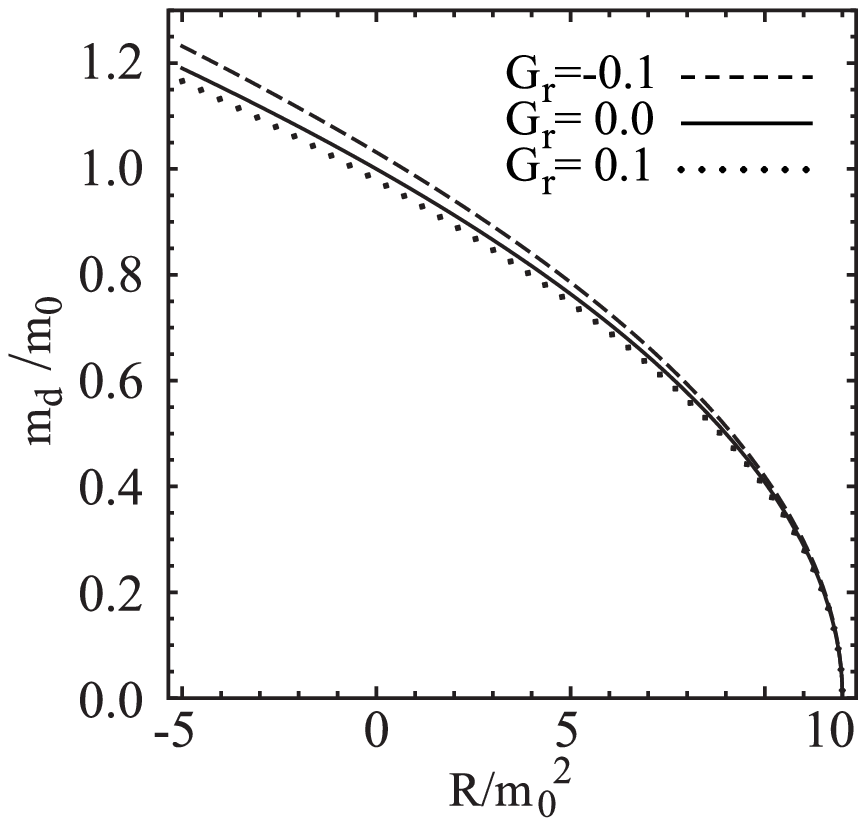}
  \end{center}
  \caption{Dynamically generated fermion mass at the four-dimensional
     limit for $G_1<0$ and $G_r=-0.1, 0, 0.1$ as a function of $R$. }
  \label{fig:6}
  \end{minipage}\hspace{0.03\hsize}
  \begin{minipage}{0.48\hsize}
    \begin{center}
      \includegraphics[width=60mm]{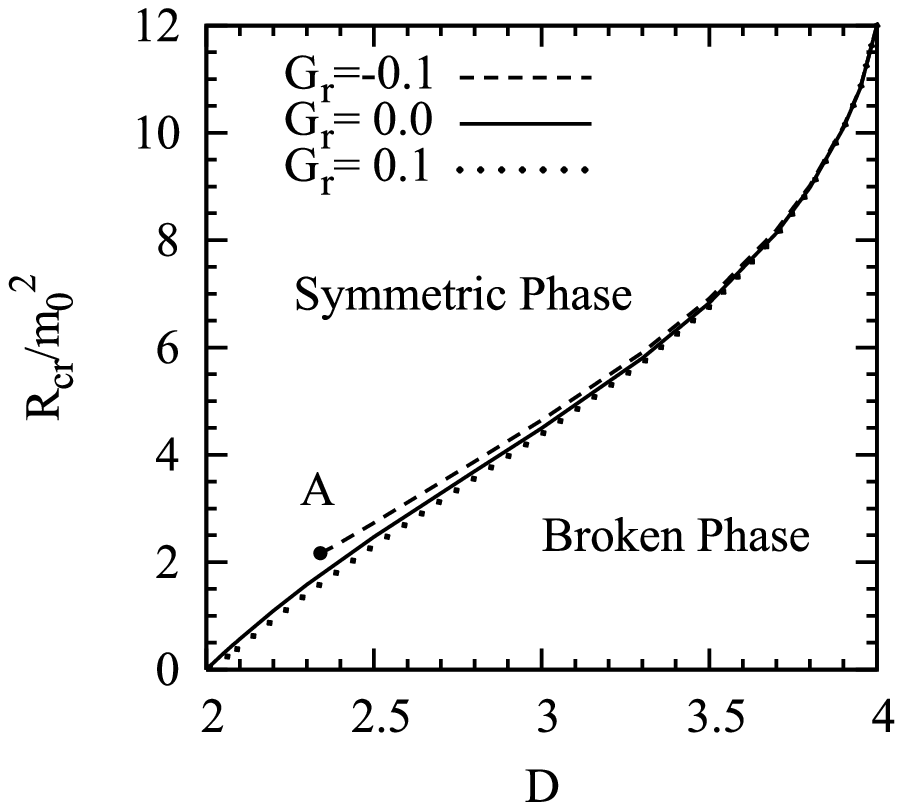}
    \end{center}
    \caption{Phase structure as a function of the space-time
      dimension $D$ for $G_1<0$, $G_r=-0.1, 0$ and $0.1$. }
    \label{fig:8}
  \end{minipage}
\end{figure}

In Figs.~\ref{fig:5} and \ref{fig:6} the dynamical fermion mass
is plotted as a function of the space-time curvature, $R$.
We numerically evaluate the true minimum of the effective 
potential, as the curvature $R$ varies,
and obtain the solution corresponding to the true minimum.
In Fig.~\ref{fig:5} the mass gap appears at the critical 
point. Thus the phase transition is of the first order for 
$2\leq D < 4$. A negative $G_r$ enhances the symmetry breaking
and a larger critical curvature is observed for a negative $G_r$.
We obtain a smaller $R_{cr}$ for a positive $G_r$.

Because of the non-renormalizability of the four- and the 
eight-fermion interactions, we can not remove all the divergence
at the four dimensional limit. For example, the right-hand side
in the gap equation (\ref{order:R}) is divergent at the limit.
If we normalize all the mass scales in the gap equation by 
$m_0$ given in Eq.(\ref{m0}), the factor $\Gamma({1-D/2})$ in
$m_0$ eliminates the one in the gap equation. Thus
we obtain a finite expression for the gap equation at the
limit, $D\rightarrow 4$,
\begin{eqnarray}
\frac{s}{m_0}
&=&-\mbox{sgn}(G_1)\frac{s}{m_0}\left[\left(\frac{s}{m_0}\right)^2
-\frac{1}{12}\frac{R}{m_0^2}\right]
\nonumber \\
&&-\mbox{sgn}(G_1)\frac{G_r}{2}\left(\frac{s}{m_0}\right)^3
\left[\left(\frac{s}{m_0}\right)^2
-\frac{1}{12}\frac{R}{m_0^2}\right]^3 .
\label{gap:4D}
\end{eqnarray}
We plot the solution of Eq.(\ref{gap:4D}) in Fig.~\ref{fig:6} for
a negative $G_1$. The broken chiral symmetry is restored through the 
second order phase transition at $R/m_0^2=12$. The critical
value is independent on the eight-fermion coupling $G_2$.
Therefore we obtain the critical curvature, $R_{cr}$, as a function
of the space-time dimension, $D$. In Fig. \ref{fig:8} we
illustrate the phase boundary which divides the symmetric 
phase and the broken phase for $G_1<0$ and $G_r=-0.1, 0, 0.1$.
We find a positiveness of the critical curvature.
Only the broken phase is realized in a space-time 
with a negative curvature.\cite{Gorbar:1999wa}
The eight-fermion interaction contributes the curvature induced
phase transition.
Especially, we observe a endpoint, A, for $G_r=-0.1$.\cite{InagakiHayashi} 
At the 
point a the local minimum disappears from the range, 
$\langle \sigma \rangle /m_0 \lesssim 1$, 
considered here. 

\subsection{Phase structure for a positive $G_1$}
The chiral symmetry is preserved in Minkowski 
space-time for a positive $G_1$.
\begin{figure}[bp]
  \begin{minipage}{0.48\hsize}
    \begin{center}
      \includegraphics[width=60mm]{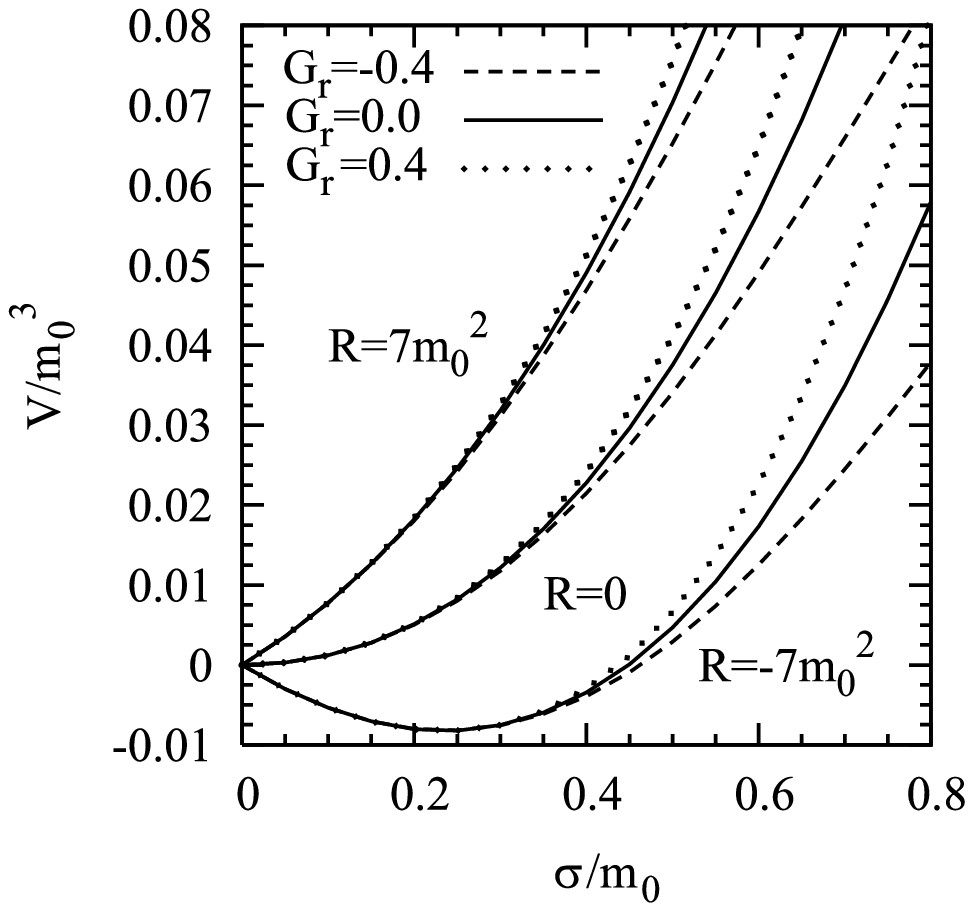}
    \end{center}
    \caption{Behavior of the effective potential for $D=3$,
     $G_1>0$ and $G_r=-0.4, 0, 0.4$ as the curvature, $R$, varies. }
    \label{fig:4}
  \end{minipage}\hspace{0.03\hsize}
  \begin{minipage}{0.48\hsize}
    \begin{center}
      \includegraphics[width=58mm]{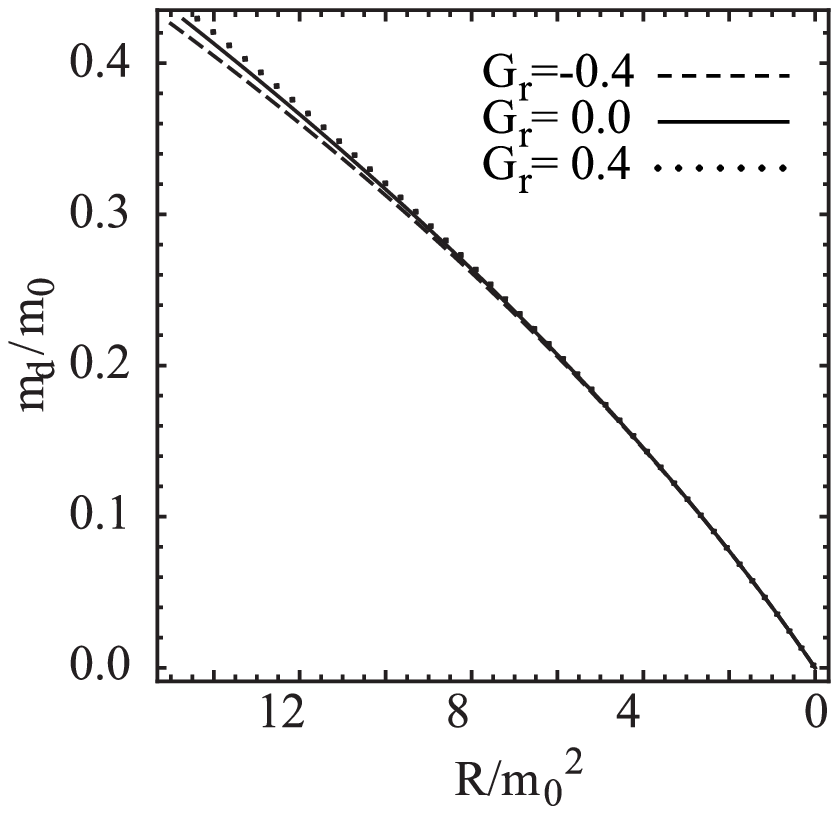}
    \end{center}
    \caption{Dynamically generated fermion mass for $D=3$,
     $G_1>0$ and $G_r=-0.4, 0, 0.4$ as a function of $R$.}
    \label{fig:9}
  \end{minipage}
\end{figure}
The space-time curvature modifies the effective 
potential, as is shown in Fig.~\ref{fig:4}.
The effective potential has a single and a double well
shape in a space-time with a non-negative and a negative 
curvature, respectively. An influence of the eight-fermion 
interaction is presented by dashed and dotted lines. 

\begin{figure}[tbp]
  \begin{minipage}{0.48\hsize}
    \begin{center}
      \includegraphics[width=60mm]{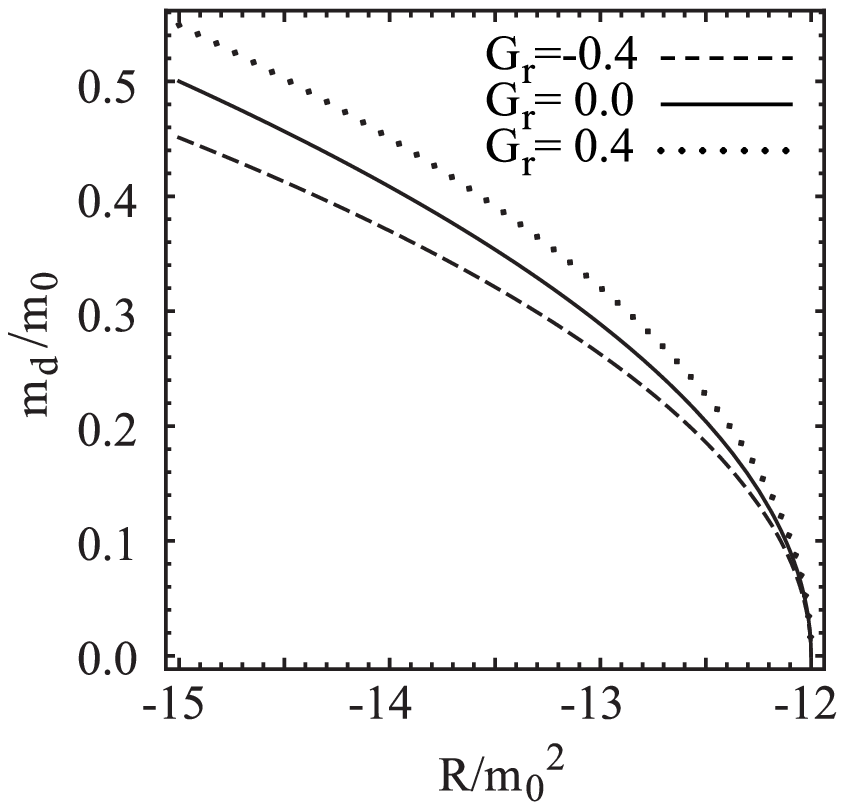}
    \end{center}
    \caption{Dynamically generated fermion mass at the four-dimensional
     limit for $G_1>0$ and $G_r=-0.4, 0, 0.4$ as a function of $R$.}
    \label{fig:10}
  \end{minipage}\hspace{0.03\hsize}
  \begin{minipage}{0.48\hsize}
    \begin{center}
      \includegraphics[width=60mm]{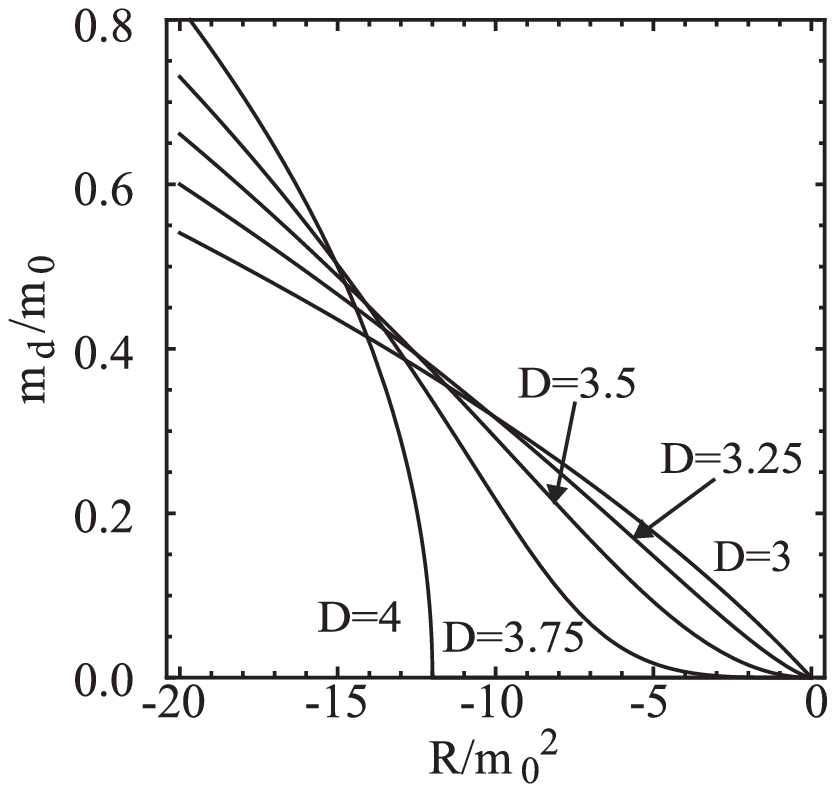}
    \end{center}
    \caption{Dynamically generated fermion mass for $G_1>0$, $G_r= 0$,
             $D=3, 3.25, 3.5, 3.75$ and the limit $D\rightarrow 4$
             as a function of $R$.}
    \label{fig:11}
  \end{minipage}
\end{figure}

Though the eight-fermion interaction modifies the effective 
potential, it can not change the phase structure for a 
positive $G_1$. Such a situation can be clearly seen on the 
behavior of the dynamical fermion mass.
We solve the gap equation (\ref{order:R}) and draw the 
dynamical fermion mass as a function of the space-time
curvature in Figs.~\ref{fig:9} and \ref{fig:10}.
As is shown in Fig.~\ref{fig:9}, the dynamically generated 
fermion mass disappears at $R=0$ for $2\leq D<4$.
Thus the critical curvature is given by $R_{cr}=0$
and the chiral symmetry is always broken for a 
negative $R$. Only the symmetric phase is realized for a 
non-negative $R$. 
In Figs.~\ref{fig:10} and \ref{fig:11} we observe that the 
critical curvature shifts to a smaller value $R_{cr}/m_0^2=-12$ 
at the four-dimensional limit.
Higher than the second order phase transition 
occurs at $R=0$ for $2\leq D<4$, while the second order phase 
transition takes place at $R/m_0^2=-12$ for $D\rightarrow 4$.
As is discussed in the previous section, the critical value is 
independent of the eight-fermion coupling.
Fig.\ref{fig:11} shows the transition from $R_{cr}=0$ to 
$R_{cr}/m_0^2=-12$ for $G_r=0$, as the space-time 
dimension, $D$, varies.

\section{Finite Size and Topological Effects}
The very early universe may have a non-trivial topology.
A symmetry restoration at high temperature can be also 
classified as the similar type of effects.
In this section we investigate the contribution from the 
eight-fermion interaction to the dynamical symmetry 
breaking in a flat space-time with non-trivial topology, 
$R^{D-1}\otimes S^1$.

In the compact manifold $S^1$ fermion fields should obey
the boundary condition,
\begin{equation}
\psi(x+L)=e^{i\delta_{p,1}\pi}\psi(x),
\label{bc}
\end{equation}
where $L$ is the size of the $S^1$ direction.
Since the boundary condition (\ref{bc}) restricts the
allowed mode functions, the Fourier transformation in 
Eq.~(\ref{eq:2p2}) is replaced by the Fourier series 
expansion,
\begin{equation}
\left\{
\begin{array}{rcl}
\displaystyle
\int^{\infty}_{\infty}\frac{dk^i}{2\pi} & \rightarrow &
\displaystyle \frac{1}{L}\sum_{n=-\infty}^{\infty},
\\
k^i & \rightarrow & \displaystyle \frac{(2n+\delta_{p,1})\pi}{L},
\end{array}
\right.
\end{equation}
where the upper index $i$ shows the compact direction, $S^1$, 
and $\delta_{p,1}$ is fixed by the boundary condition.
Thus the spinor two point function (\ref{eq:2p}) in 
$R^{D-1}\otimes S^1$ is found to be
\begin{equation}
  S(x,y;s)=\frac{1}{L}\sum_{n=-\infty}^{\infty}
  \int \frac{d^{D-1} k}{(2\pi)^{D-1}}e^{-ik(x-y)}
  \frac{1}{\gamma^\mu k_\mu -s}.
\label{2p:S1}
\end{equation}
Substituting Eq.~(\ref{2p:S1}) into Eq.~(\ref{pot}),
we obtain the effective potential in $R^{D-1}\otimes S^1$,
\begin{eqnarray}
V(s,\sigma')&=&
-\frac{\sigma'^2}{4G_1}+\frac{s^2}{4G_1}
-\frac{G_2}{16G_1^4}(\sigma'+s)^4
+\frac{\mbox{tr1}}{2(4\pi)^{(D-1)/2}}
\Gamma\left(\frac{1-D}{2}\right)
\frac{1}{L}
\nonumber\\
&&\times\sum_{n=-\infty}^{\infty}
\left[\left(\frac{(2n+\delta_{p,1})\pi}{L}\right)^2
+s^2\right]^{(D-1)/2}.
\label{pot:L}
\end{eqnarray}
The stationary condition (\ref{gap:2}) is not modified. 
The auxiliary fields $s$ and $\sigma'$ are represented
by $\sigma$ at the stationary point.
We insert Eq.~(\ref{gapsol:1}) into Eq.~(\ref{pot:L}) 
and rewrite the effective potential,
\begin{eqnarray}
V(\sigma)&=&
\frac{1}{4G_1}\left(\sigma^2+
\frac{3G_2}{4G_1^3}\sigma^4\right)
+\frac{\mbox{tr1}}{2(4\pi)^{(D-1)/2}}
\Gamma\left(\frac{1-D}{2}\right)
\frac{1}{L}
\nonumber\\
&&\times\sum_{n=-\infty}^{\infty}
\left[\left(\frac{(2n+\delta_{p,1})\pi}{L}\right)^2
+\left(
\sigma+
\frac{G_2}{2G_1^3}\sigma^3
\right)^2
\right]^{(D-1)/2}. 
\label{pot:L2}
\end{eqnarray}
The necessary condition for the dynamical fermion mass is given 
by the gap equation defined by (\ref{gapsol:2}). 
We substitute the two-point function (\ref{2p:S1}) into 
Eq.~(\ref{gapsol:2}) and obtain
\begin{eqnarray}
s&=&2G_1\frac{\mbox{tr}1}{(4\pi)^{(D-1)/2}}
\Gamma\left(\frac{1-D}{2}\right)\frac{s}{L}\sum_{n=-\infty}^{\infty}
\left[\left(\frac{(2n+\delta_{p,1})\pi}{L}\right)^2
+s^2\right]^{(D-3)/2}
\nonumber \\
&&+4G_2\Biggl\{
\frac{\mbox{tr}1}{(4\pi)^{(D-1)/2}}
\Gamma\left(\frac{1-D}{2}\right)\frac{s}{L}
\nonumber \\
&&\times\sum_{n=-\infty}^{\infty}
\left[\left(\frac{(2n+\delta_{p,1})\pi}{L}\right)^2
+s^2\right]^{(D-3)/2}
\Biggl\}^3 .
\label{gap:top}
\end{eqnarray}
Below we evaluate the effective potential (\ref{pot:L2}) and 
the gap equation (\ref{gap:top}) for fermions with the 
periodic, $\delta_{p,1}=0$, and the anti-periodic 
boundary conditions, $\delta_{p,1}=1$ and discuss the phase
structure.

\subsection{Phase structure for a negative $G_1$}
In Minkowski space-time, $R^D$, the system is in the 
broken phase for a negative $G_1$. The space-time, $R^D$, 
is obtained at the large $L$ limit, $L\rightarrow\infty$, 
of $R^{D-1}\otimes S^1$. We start from the broken phase
at the large $L$ limit and evaluate the effective
potential as $L$ decreases.

\begin{figure}[bp]
  \begin{minipage}{0.48\hsize}
    \begin{center}
      \includegraphics[width=60mm]{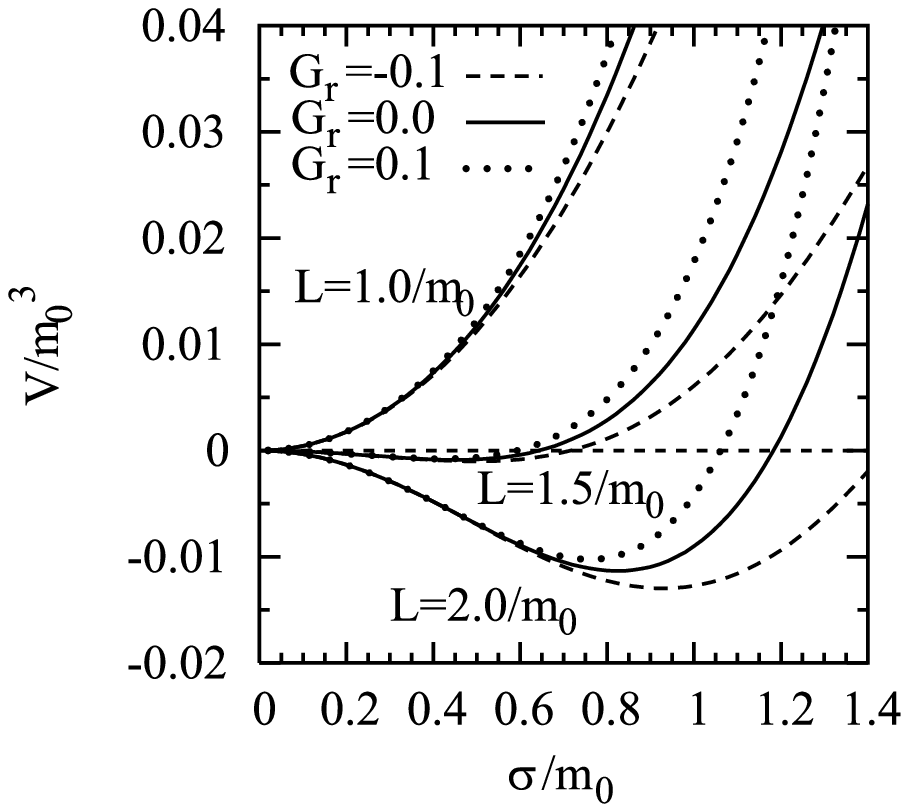}
    \end{center}
      \caption{Behavior of the effective potential for
         $D=3$, $G_1<0$ and $G_r=-0.1, 0, 0.1$
	in the case of the anti-periodic boundary condition.}
      \label{fig:13}
  \end{minipage}\hspace{0.03\hsize}
  \begin{minipage}{0.48\hsize}
    \begin{center}
      \includegraphics[width=60mm]{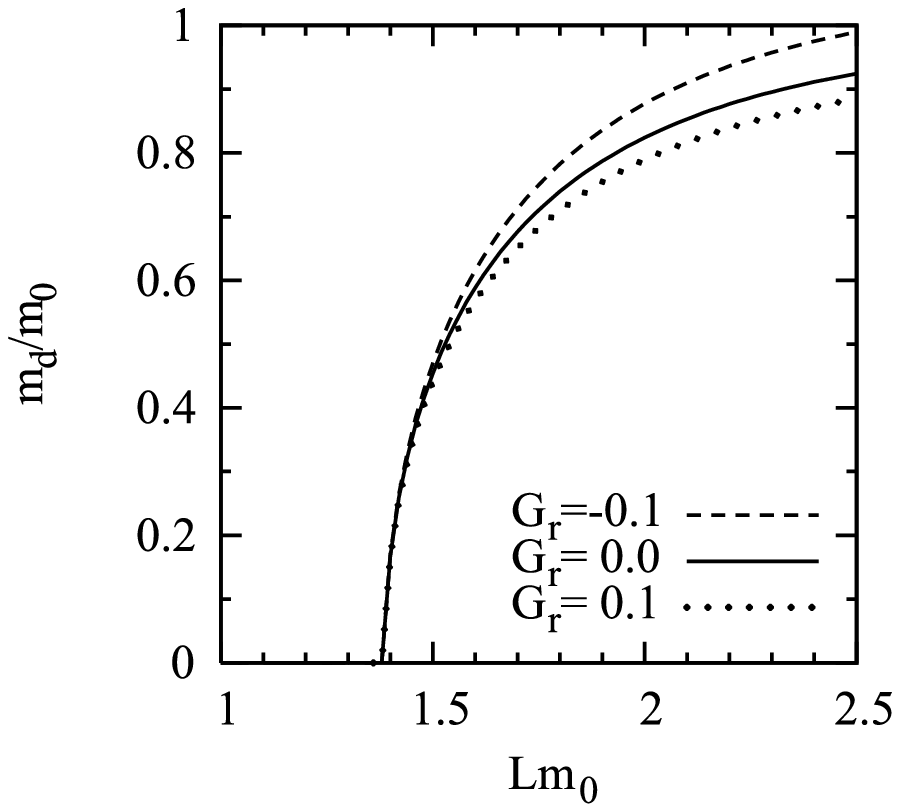}
    \end{center}
    \caption{Dynamical fermion mass for $D=3$,
         $G_1<0$ and $G_r=-0.1, 0, 0.1$ as a function of
         the length $L$
	in the case of the anti-periodic boundary condition.}
\label{fig:14}
  \end{minipage}
\end{figure}

In the case of the anti-periodic boundary condition the 
theory on $R^{D-1}\otimes S^1$ is equivalent to the finite 
temperature theory. It is expected that a broken symmetry 
is restored at high temperature. Thus the finite size 
effect should also restore the broken symmetry.
We draw the typical behavior of the effective potential
(\ref{pot:L2}) for the fermion field with the anti-periodic 
boundary condition in Fig.~\ref{fig:13}. The broken chiral 
symmetry is restored for a small $L$. 

\begin{figure}[tbp]
  \begin{minipage}{0.48\hsize}
    \begin{center}
      \includegraphics[width=65mm]{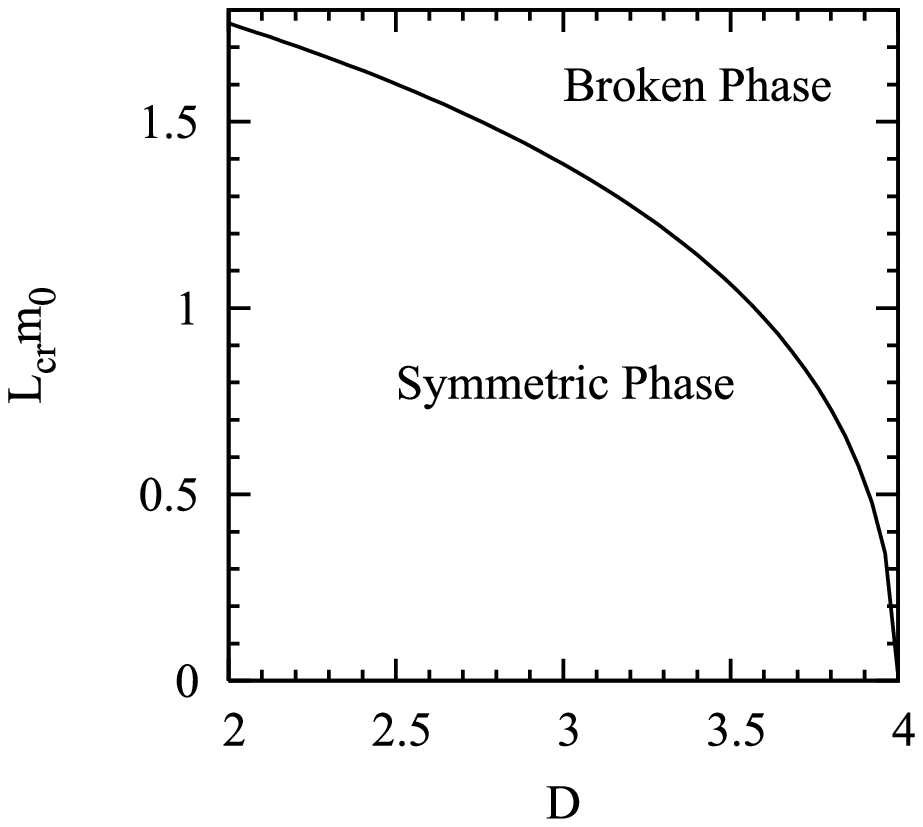}
    \end{center}
    \caption{Phase diagram for $G_1<0$ in the case of the 
              anti-periodic boundary condition.
      \label{fig:20}}
  \end{minipage}\hspace{0.03\hsize}
  \begin{minipage}{0.48\hsize}
    \begin{center}
      \includegraphics[width=60mm]{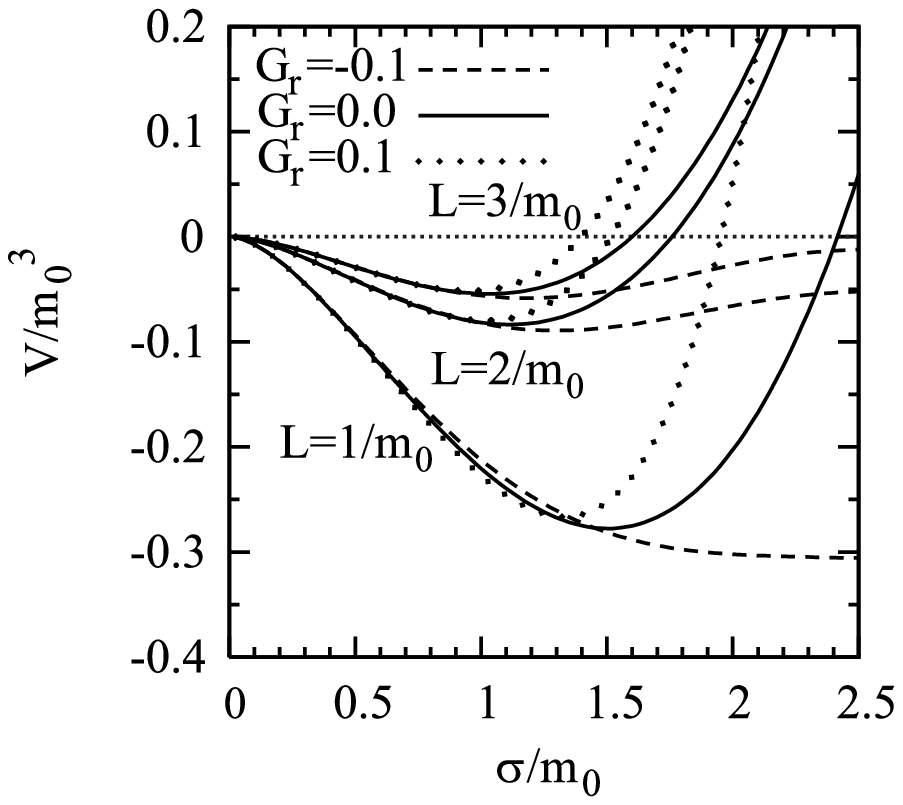}
    \end{center}
     \caption{Behavior of the effective potential for
         $D=3$, $G_1<0$ and $G_r=-0.1, 0, 0.1$
	in the case of the periodic boundary condition.}
      \label{fig:15}
  \end{minipage}
\end{figure}

The dynamically generated mass for the fermion field with 
the anti-periodic boundary condition is 
plotted in Fig.~\ref{fig:14} by solving the gap equation
(\ref{gap:top}). It is clearly seen that the dynamical
fermion mass disappears and the broken chiral symmetry is 
restored through the second order phase transition at the 
critical length, $L_{cr}$.
The dynamical fermion mass is modified by the eight-fermion
interaction. As is discussed in Sec.~2, the eight-fermion 
interaction does not affect the critical length, $L_{cr}$.
We can analytically calculate the critical length, $L_{cr}$,
by taking the limit $s \rightarrow 0$ for the non-trivial 
solution of the gap equation (\ref{gap:top}). Thus we
find the explicit expression for the critical length, $L_{cr}$,
\begin{equation}
L_{cr}=
2\pi\left[\frac{2\mbox{tr1}G_1}{\pi(4\pi)^{(D-1)/2}}
(2^{3-D}-1)
\Gamma\left(\frac{3-D}{2}\right)\zeta(3-D)\right]^{1/(D-2)}.
\end{equation}
for fermion fields with the anti-periodic boundary condition.
We plot it as a function of the space-time dimension, $D$, 
in Fig.~\ref{fig:20}. Though the effective potential 
(\ref{pot:M}) is divergent at the four-dimensional limit,
the finite size effect gives a finite correction for the 
divergent potential. The effect disappears at the 
four-dimensional limit.

\begin{figure}
  \begin{minipage}{0.48\hsize}
    \begin{center}
      \includegraphics[width=65mm]{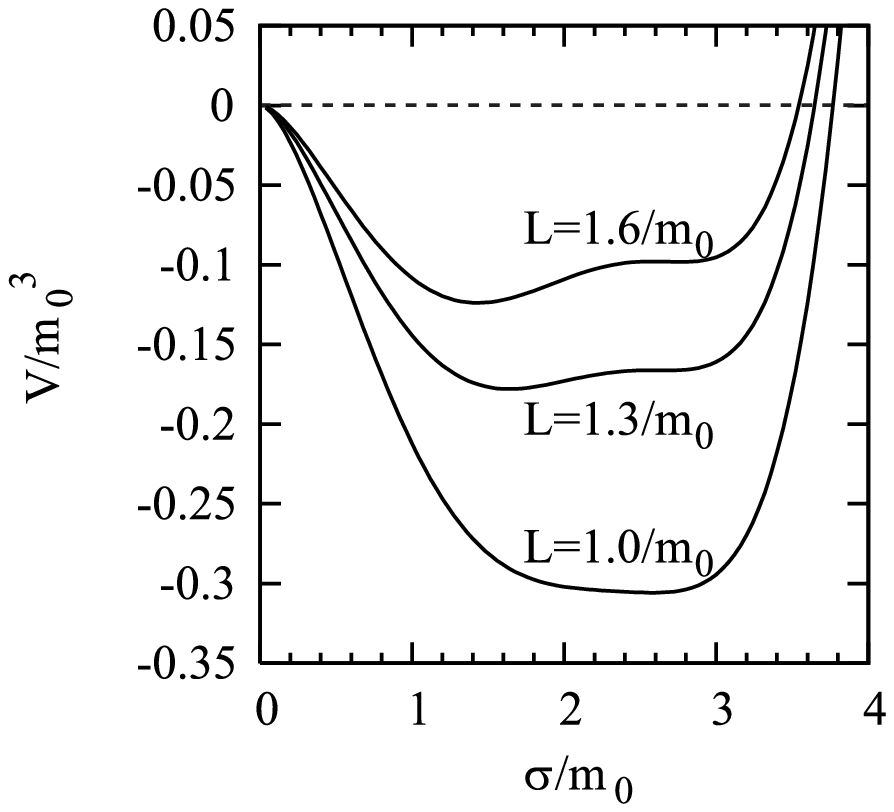}
    \end{center}
    \caption{Behavior of the effective potential for
         $D=3$, $G_1<0$ and $G_r=-0.1$
	in the case of the periodic boundary condition.}
  \label{fig:16}
  \end{minipage}\hspace{0.03\hsize}
  \begin{minipage}{0.48\hsize}
    \begin{center}
      \includegraphics[width=60mm]{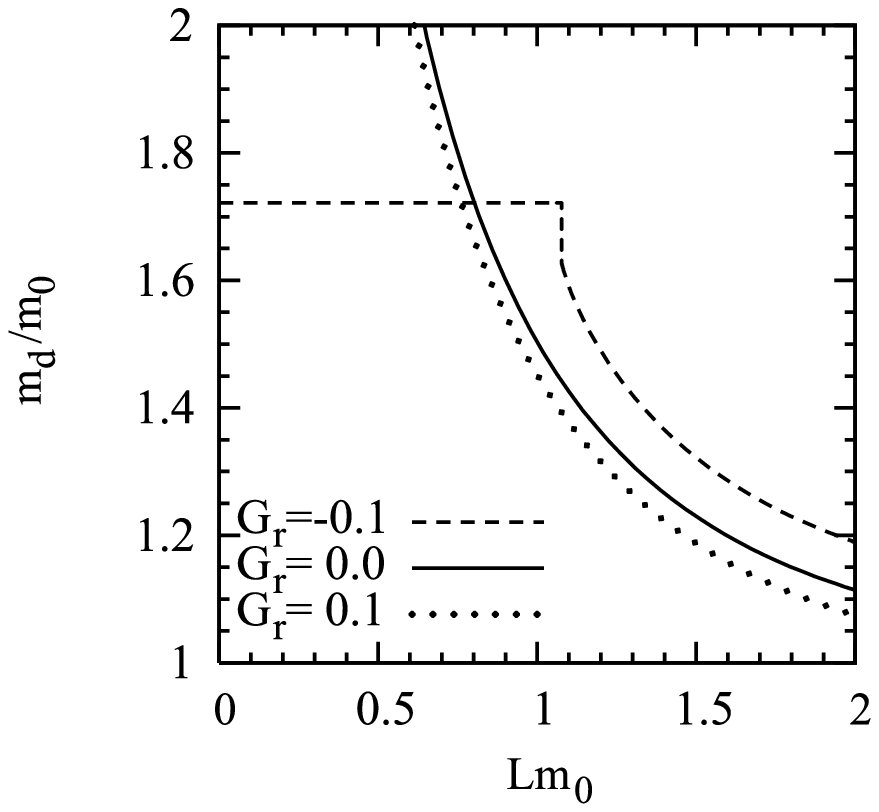}
    \end{center}
    \caption{Dynamical fermion mass for $D=3$,
         $G_1<0$ and $G_r=-0.1, 0, 0.1$ as a function of
         the length $L$
	in the case of the periodic boundary condition.}
  \label{fig:17}
  \end{minipage}
\end{figure}

We draw the behavior of the effective potential
for the fermion field with the periodic boundary condition
in Fig. \ref{fig:15}. The finite size effect enhances 
the chiral symmetry breaking in this case. Thus only the 
broken phase can be realised for a negative $G_1$. If we 
enlarge the scope of our analysis to include a larger 
$\sigma$, the eight-fermion interaction induces 
another type of transition between finite $\sigma$. 
As is shown in Fig. \ref{fig:16}, the effective potential 
can develop two local minima for $G_r=-0.1$. The finite
size effect shifts the true minimum from the first to 
the second local one which is outside the scope of our
interest.
In Fig. \ref{fig:17} we plot the dynamical fermion mass in 
the case of the periodic boundary condition by solving the
gap equation (\ref{gap:top}). The finite size effect increases 
the dynamically generated fermion mass.
For a negative $G_r$ we observe a small mass gap which is induced
by the transition from the first to the second minimum in Fig.~15.
Then the two minima combined at $\sigma/m_0=\sqrt{-2/(3G_r)}$, as
$L$ deceases. If the size $L$ is small enough, the dynamical
fermion mass is given by
\begin{equation}
m_d/m_0=\sqrt{-\frac{2}{3G_r}}+\frac{G_r}{2}
\left(-\frac{2}{3G_r}\right)^{3/2},
\label{mass:fixed}
\end{equation}
for a negative $G_r$.

\subsection{Phase structure for a positive $G_1$}
For a positive $G_1$ the system is in the symmetric phase
at the Minkowski limit, $L\rightarrow \infty$. To study the 
chiral symmetry breaking induced by the finite size effect
we evaluate the effective potential (\ref{pot:L2}).
In the case of the anti-periodic boundary condition the 
finite size effect works to stabilize the trivial ground
state at $\sigma=0$. Thus only the symmetric phase 
can be realized. We can not observe any transition of the 
ground state. Therefore it is enough to analyze only
fermion fields which obey the periodic boundary condition.

\begin{figure}[tbp]
  \begin{minipage}{0.48\hsize}
    \begin{center}
      \includegraphics[width=63mm,height=57mm]{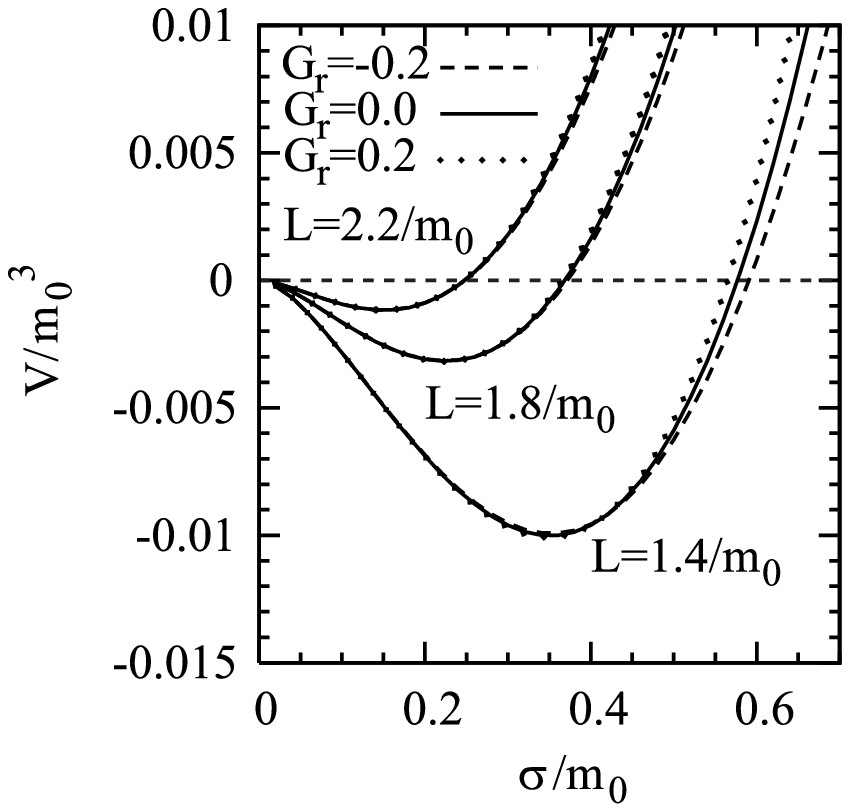}
    \end{center}
      \caption{Behavior of the effective potential for
         $D=3$, $G_1>0$ and $G_r=-0.2, 0, 0.2$
	in the case of the periodic boundary condition.}
      \label{fig:18}
  \end{minipage}\hspace{0.03\hsize}
  \begin{minipage}{0.48\hsize}
    \begin{center}
      \includegraphics[width=60mm]{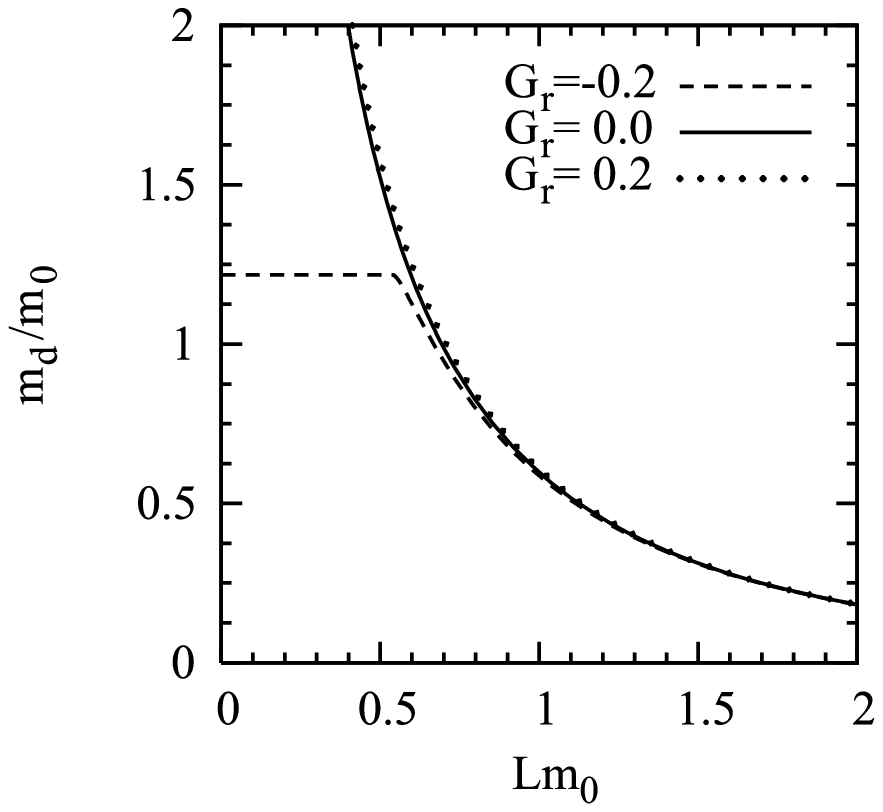}
    \end{center}
    \caption{Dynamical fermion mass for $D=3$,
         $G_1>0$ and $G_r=-0.2, 0, 0.2$ as a function of
         the length $L$
	in the case of the periodic boundary condition.}
  \label{fig:19}
  \end{minipage}
\end{figure}
\begin{figure}[tbp]
  \begin{minipage}{0.48\hsize}
    \begin{center}
      \includegraphics[width=55mm]{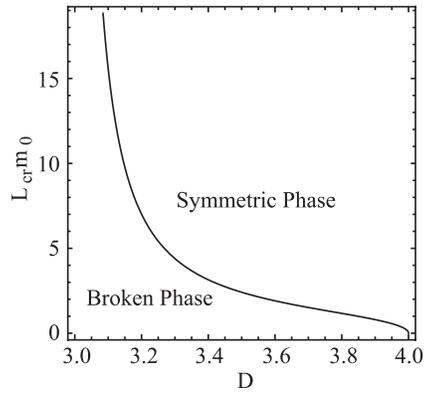}
    \end{center}
    \caption{Phase diagram for $G_1>0$ in the case of the 
              periodic boundary condition.}
  \label{fig:21}
  \end{minipage}
\end{figure}

For fermion fields with the periodic boundary condition
we plot the behavior of the effective potential in 
Fig.~\ref{fig:18}. The effective potential has the double 
well shape for a small $L$. Thus the phase transition is
caused by the finite size effect. The order of the transition
is found by observing the dynamical fermion mass.
We draw it as a function of $L$ in Fig. \ref{fig:19}. 
It is found that the phase transition is of higher than 
the second order.
The eight-fermion interaction has larger contribution
for a negative $G_r$. In this case the dynamical fermion
mass is give by Eq.(\ref{mass:fixed}) again, if $L$ is
small enough.

The critical length, $L_{cr}$, is derived by taking the 
limit $s \rightarrow 0$ for the non-trivial solution of 
the gap equation (\ref{gap:top}). It is found to be
\begin{equation}
L_{cr}=
2\pi\left[\frac{2\mbox{tr1}G_1}{\pi(4\pi)^{(D-1)/2}}
\Gamma\left(\frac{3-D}{2}\right)\zeta(3-D)\right]^{1/(D-2)},
\end{equation}
for fermion fields with the periodic boundary condition.
The chiral symmetry is broken at $L=L_{cr}$.
We illustrate the phase diagram for $G_1>0$ in the case of the
periodic boundary condition in Fig. \ref{fig:21}. 
The eight-fermion interaction has nothing to do with the phase 
boundary. It should be noted that the finite size effect 
disappears at the four-dimensional limit.

\section{Conclusion}
In this article the four- and eight-fermion interaction model has been 
investigated in the two types of space-time, a weakly curved space-time 
and a cylindrical space-time, $R^{D-1}\otimes S^1$. 
The expectation value for the composite operator,
$\bar{\psi}\psi$ is one of the order parameters for the 
chiral symmetry breaking. Thus the phase structure of the system is 
found by observing the effective potential in terms of the order 
parameter, $\sigma (\propto \bar{\psi}\psi)$.
We have applied the dimensional regularization and have calculated 
the effective potential in the leading order of the $1/N$ expansion.
The effective potential is written as a function of the auxiliary
field, $\sigma$, space-time dimension, $D$, a ration of the coupling
constant, $G_r=G_2/G_1^3$, and the sign of the four-fermion
coupling, sgn$(G_1)$. 
It has been shown that the stability of the critical point against 
the eight-fermion interaction. The eight-fermion interaction 
can not modify the phase boundary, if the phase transition is of the 
second or higher than second order. 

In Minkowski space-time, $R^D$, the chiral symmetry is broken for 
a negative four-fermion coupling. As is mentioned in 
Ref.~\cite{IKK}, the renormalized coupling can be positive.
The eight-fermion interaction modifies the shape of the effective 
potential. Contributions from the eight-fermion interaction
is enhanced for a larger $\sigma (\propto \bar{\psi}\psi)$. 
However, if the auxiliary field $\sigma$ develops a large value,
we can not neglect higher dimensional operators which are not 
employed. We have analysed the system in the restricted parameter
range, $\langle \sigma \rangle /m_0 \lesssim 1$. 

To study the curvature effect we have evaluated the effective 
potential in weakly curved space-time. In the four-fermion 
interaction model the broken symmetry is restored for a large 
positive curvature, $R$. The phase transition is of the first
order for a negative four-fermion coupling, $G_1$ at $2\leq D <4$.
Only the broken phase appears in a space-time with a negative 
curvature. The dynamically generated fermion mass is modified
by the eight-fermion interaction. Except for the four-dimensional
limit we have observed a lower and a higher critical curvature 
for a positive and a negative $G_r$, respectively. The phase
boundary is shown as a function of $D$.
If we consider a negative $G_2$, the minimum of the effective 
potential disappears inside the range, 
$\langle \sigma \rangle /m_0 \lesssim 1$.
The phase boundary vanishes at a low dimension for $G_2<0$.
In Ref.~\cite{HIT} the cutoff regularization is applied
to analyse the same model in weakly curved space-time at $D=4$.
Since the four- and eight-fermion interaction are 
non-renormalizable in four dimensions, the phase structure 
depends on the regularization procedures.

The space-time topology also has non-negligible contribution
for the expectation value of the composite operator, 
$\bar{\psi}\psi$. We have investigated the flat space-time 
with nontrivial topology, $R^{D-1}\otimes S^{-1}$ and found 
the phase structure of the system. The finite size effect can 
induce the second or higher than second order phase transition. 
Thus the critical length does not modified by the eight-fermion 
interaction. We have obtained the phase boundary independent
of the eight-fermion coupling, $G_2$.
The system for fermion fields with the anti-periodic boundary 
condition is equivalent to the finite temperature field theory
in Matsubara Formalism. There is a correspondence between the
phase boundary in Fig.~\ref{fig:20} and the one at finite
temperature, $T$, if we make a replacement,
\begin{equation}
L \leftrightarrow \frac{1}{k_B T},
\end{equation}
For a negative $G_r$ and $G_1$ two local minima are observed in the 
effective potential, though the second one is outside the range, 
$\langle \sigma \rangle /m_0 \lesssim 1$.

In the present paper we have studied a simple toy model of the
dynamical symmetry breaking. But we believe that the curvature 
and the topological effects on the dynamical symmetry breaking
is similar in general to the case we have investigated.
To distinguish a characteristic feature of the model we
will continue our works in other models, vector-type 
fermion interactions, gauge theories and the supersymmetric
extension of the model and so on in various background 
space-time.
Here we focus on the mathematical aspects of the four- and 
eight-fermion interaction model. However, we are interested 
in the application of our analysis to phenomena at the early 
universe. It is expected that the spontaneous symmetry breaking
has play an essential role for the evolution of the space-time.
To study the evolution of the universe we should calculate
the stress tensor and solve the Einstein equation. We will 
continue our works and hope to report on these problems.

\section*{Acknowledgment}
The authors would like to thank H.~Takata, D.~Kimura, Y.~Kitadono and
 Y.~Mizutani for fruitful discussions.
T.~I. is supported by the Ministry of Education, Science, Sports and
Culture, Grant-in-Aid for Scientific Research (C), No. 18540276, 2009.


\end{document}